\documentstyle[titlepage,12pt]{article}
\begin{document}
\renewcommand{\theequation}{\thesection.\arabic{equation}}
\title{New Coherent String States and Minimal Uncertainty in WZWN Models}
\author{ A.L. Larsen\thanks{Department of
Physics, University of Odense,
Campusvej 55, 5230 Odense M, Denmark.}
and
N. S\'{a}nchez\thanks{Observatoire de Paris,
DEMIRM. Laboratoire Associ\'{e} au CNRS, Observatoire de Paris et
\'{E}cole Normale Sup\'{e}rieure. 61, Avenue
de l'Observatoire, 75014 Paris, France.}}
\maketitle
\begin{abstract}
We study the properties of {\bf exact} (all level $k$) quantum coherent
states in the context of string theory on a group manifold (WZWN models).
Coherent states of WZWN models may help to solve the unitarity problem:
Having positive norm, they consistently describe the very massive string
states (otherwise excluded by the spin-level condition). These states can be
constructed by  (at least) two alternative procedures: (i) as the
exponential of
the creation operator on the ground state, and (ii) as eigenstates of the
annhilation operator. In the $ k \to \infty $    limit, all the known
properties of
ordinary coherent states are recovered. States (i) and (ii) (which are
equivalent
in the context of ordinary quantum mechanics and string theory in flat
spacetime)
are not equivalent in the context of WZWN  models. The set (i) was
constructed by
these authors in a previous article. In  this paper we provide the
construction of
states (ii), we compare the two sets and discuss their properties. We analyze
the uncertainty
relation,  and show that states (ii) satisfy automatically the {\it minimal
uncertainty} condition for any $k$; they are thus  {\it
quasiclassical}, in some
sense more classical than states (i) which only satisfy it in the $
k \to \infty $ limit. Modification to the Heisenberg relation is given by
$ 2\,{\cal H }/k $, where ${\cal H }$ is connected to the string energy.
\end{abstract}
\newpage

\section{Introduction}

\setcounter{equation}{0}
Coherent states play an important role in quantum mechanics, where they
represent the "quasi-classical" states of minimal uncertainty (see for instance
\cite{merz}). Coherent states have also been widely used in string theory
for the
computation of scattering amplitudes (see for instance \cite{green}).  Both in
ordinary quantum mechanics and in the standard formulation of string theory,
coherent states can be defined in (at least) two alternative but {\it
equivalent}
ways: Either as eigenstates of the annihilation operator, or as the
exponential of
the creation operator acting on the ground state. However, when considering
string
theory on a group manifold using a WZWN or gauged WZWN construction, the
situation
is quite different. In that case, the fundamental Abelian harmonic oscillator
commutator $[a,a^\dagger]=1$ is substituted by a non-Abelian Kac-Moody current
algebra. As a consequence, a state defined as the exponential of a creation
operator acting on the ground state, will no longer in general be an
eigenstate of
the annihilation operator. For the same reason, it will generally not be a
state of
minimal uncertainty.

The purpose of the present paper is to discuss the alternative, but
{\it in-equivalent}, definitions of coherent states in WZWN models. More
precisely, we shall consider the $SL(2,R)$ WZWN model corresponding to  bosonic
string theory in 3-dimensional Anti de Sitter space, $AdS_3\cong\;
SL(2,R)\cong\;
SU(1,1) $. It represents the simplest example of string theory on a
manifold with
curved space {\it and} curved time [3-16], and it has attracted renewed
interest
recently in the context of the  conjecture \cite{malda} connecting
supergravity and
superstring theory in $AdS$ space with a conformal field theory on the
boundary.
For a recent discussion about coherent states on group manifolds, see
also \cite{sen}.

In a previous publication \cite{all}, we considered coherent states in $AdS_3$
using the definition corresponding to the exponential of a creation
operator. Such states were shown to describe, among other things, the very
massive
string states in $AdS_3$. In particular, it was shown that there is a discrete
spectrum of very massive string states, with asymptotic behaviour
$m^2\alpha'\propto N^2$ ($N$ positive integer). This was in precise
agreement with the previous results obtained using semi-classical
quantization \cite{san2,vega1,vega}, and the same asymptotic behaviour was also
obtained in ref.\cite{Ooguri}, although the construction there was
completely different from ours.

The coherent states, defined in terms of the exponential of a creation
operator,
however, are not eigenstates of the annihilation operator, and they are not
minimal
uncertainty states. Moreover, they are somewhat complicated to work with.
In this paper we consider the alternative definition of coherent states in
$AdS_3$,
taking the property of being an eigenstate of the annihilation operator as the
fundamental one. Such states will also be automatically states of minimal
uncertainty, and they are generally much easier to work with.

There is an extensive literature on coherent states of various kinds; canonical
coherent states, spin coherent states, group-realated coherent states etc.
(for a
review see \cite{klauder}). The coherent states constructed in this paper
generalize the $SU(1,1)$ group-related coherent states originally
constructed in
\cite{barut}. The coherent states in \cite{barut} were constructed using the
ladder-operators. The ladder-operators correspond to the zero-modes of the
Kac-Moody
algebra: $L^{\pm}=\frac{1}{\sqrt{2}}J_0^\pm$, $L_{12}=J_0^3$. In string
theory, the
zero-modes are not really creation and annihilation operators. Therefore,
we use
the
$n=1$ modes instead. So our coherent states are completely different from
theirs.
However, it is interesting that our results reduce to theirs if we take
$k=0$, where
$k$ is the level of the WZWN model. The reason is that the $n=1$ algebra, for
$k=0$, is formally identical to the zero-mode algebra. Thus, all our
results for
$k=0$ reduce to theirs.

This paper is organized as follows. In Section 2, we first review the standard
formulation of string theory on a group manifold \cite{wit,wit2}. We then
derive the explicit expression for the coherent states defined as
eigenstates of the
annihilation operator. Normalization, Virasoro and mass-shell conditions are
also considered. In Section 3, we show that the coherent states constructed in
Section 2 are states of minimal uncertainty, in the usual sense of ordinary
quantum mechanics. In Section 4, we consider the relation between these "new"
coherent states and the "old" coherent states discussed in \cite{all}.  Finally
in Section 5, we have some concluding remarks.

\section{Coherent States}
\setcounter{equation}{0}

The $SL(2,R)$ Kac-Moody algebra for (say) the left-moving currents is
given by
\begin{equation}
[J^a_m,J^b_n]=i\epsilon^{ab}\;_c J^c_{m+n}+\frac{k}{2}m\eta^{ab}\delta_{n+m}
\end{equation}
where
$$
k = {1 \over H^2 \; \alpha' } \; ,
$$
is the level of the $SL(2,R)$ WZWN model, and $ H^{-1} $ stands for
the length scale. (Our conventions are: $\eta^{ab}={\mbox{diag}}(1,1,-1)$ and
$\epsilon^{123}=+1$).

In terms of the currents, $J^\pm=J^1\pm iJ^2$, the algebra becomes
\begin{eqnarray}
&[J^+_m,J^-_n]&= -2J^3_{m+n}+km\delta_{m+n}\nonumber\\
&[J^3_m,J^{\pm}_n]&=\pm J^{\pm}_{m+n}\\
&[J^3_m,J^3_n]&=-\frac{k}{2}m\delta_{m+n}\nonumber
\end{eqnarray}
The world-sheet energy-momentum tensor takes the Sugawara form
\begin{equation}
T=\frac{1}{k-2}\;\eta_{ab}:J^a J^b:\;=\frac{1}{k-2}
:\left(J^+J^- -J^3 J^3\right):
\end{equation}
Its Fourier modes
\begin{equation}
T=\sum_{n=-\infty}^\infty\;L_n\;e^{-in\sigma}
\end{equation}
are given by
\begin{equation}
L_n=\frac{1}{k-2}\sum_{l=-\infty}^\infty\; :\left( \frac{1}{2}(J^+_{n-l}J^-_l+
J^-_{n-l}J^+_l)-J^3_{n-l}J^3_l\right):
\end{equation}
They fulfill the Virasoro algebra
\begin{equation}
[L_m,L_n]=(m-n)L_{m+n}+\frac{c}{12}m(m^2-1)\delta_{m+n}
\end{equation}
where the central charge is given by
\begin{equation}
c=\frac{3k}{k-2}
\end{equation}
Demanding $c=26$, corresponding to conformal invariance, gives $k=52/23$.
Notice also the commutators
\begin{equation}
[L_n,J^{\pm}_m]=-mJ^{\pm}_{n+m},\;\;\;\;\;\;\;\;
[L_n,J^3_m]=-mJ^3_{n+m}
\end{equation}
which will be usefull in the following.

The Kac-Moody algebra contains the subalgebra of zero modes $J^a_0$, for which
the quadratic Casimir is
\begin{equation}
Q=\eta_{ab}J^a_0 J^b_0=\frac{1}{2}\left( J^+_0 J^-_0 +J^-_0 J^+_0\right)-
J^3_0 J^3_0
\end{equation}
The primary states, which are quantum
states $|jm>$ at grade zero ("base-states" or "ground-states"), are
characterised by
\begin{equation}
Q|jm>=-j(j+1)|jm>,\;\;\;\;\;\;\;\;J^3_0|jm>=m|jm>
\end{equation}
Moreover, they fulfill
\begin{equation}
J^{\pm}_0|jm>=\sqrt{m(m\pm 1)-j(j+1)}\;|jm\pm 1>
\end{equation}
as well as
\begin{equation}
J^a_l|jm>=0;\;\;\;\;\;l>0
\end{equation}
The primary
states must belong to one of the unitary representations of $SL(2,R)$
(or its covering group) \cite{dixon,barg}.

For simplicity and clarity of the construction, we concentrate in the
following on the subalgebra generated by $(J^-_{+1}, J^+_{-1},J^3_0)$
\begin{eqnarray}
&[J^-_{+1},J^+_{-1}]&= 2J^3_{0}+k\nonumber\\
&[J^3_0,J^{+}_{-1}]&=J^{+}_{-1}\\
&[J^3_0,J^-_{+1}]&=-J^-_{+1}\nonumber
\end{eqnarray}
An example of string configurations described by this subalgebra is
provided by circular strings (which contain only modes corresponding
to $ n = 0 $  and $ n = \pm 1 $); it should be stressed that we
consider this subalgebra only for simplicity and clarity and that our
construction can be easily used for other string configurations as
well.

Moreover, our coherent states will be constructed using the base-state $|jj>$,
which belongs to the highest weight discrete series $D^-_j$
\cite{dixon,barg}, with states $|jm>$
\begin{equation}
j\leq -1/2\;,\;\;\;\;\;\;\;\;m=j,j-1,...
\end{equation}
Since we shall consider the covering group
of $SL(2,R)$, there are no further restrictions on $j$, i.e., it does
not need to be an integer or half-integer \cite{dixon,barg}.
In particular, from eq.(2.11) it follows that
\begin{equation}
J^+_0|jj>=0\; ,\;\;\;\;\;\;\;\; J^-_0|jj>=\sqrt{-2j}\;|jj-1>
\end{equation}
The idea is now to construct coherent states as eigenstates of the annihilation
operator $J^-_{+1}$
\begin{equation}
J^-_{+1}|\mu>=\mu|\mu>
\end{equation}
where $\mu$ is a complex number. For the state $|\mu>$ we use the ansatz
\begin{equation}
|\mu>={\cal{N}}\sum_{n=0}^\infty C_n\left( J^+_{-1}\right)^n |jj>
\end{equation}
where ${\cal{N}}$ is a normalization constant, and the coefficients $C_n$
are to be
determined. Using the commutator
\begin{equation}
[J^-_{+1},\left(J^+_{-1}\right)^n]=n\left(J^+_{-1}\right)^{n-1}(n-1+k+2J^3_0)
\end{equation}
eq.(2.16) immediately leads to the recursion
relation
\begin{equation}
(n+1)(n+2j+k)C_{n+1}=\mu C_n
\end{equation}
which is solved by
\begin{equation}
C_n=\mu^n\frac{\Gamma(2j+k)}{\Gamma(n+1)\Gamma(2j+k+n)}
\end{equation}
with the normalization $C_0=1$.

Using the identity
\begin{equation}
<jj|\left(J^-_{+1}\right)^n
\left(J^+_{-1}\right)^m|jj>=\delta_{nm}\frac{\Gamma(n+1)\Gamma(2j+k+n)}{\Gamma(2
j+k)}
\end{equation}
the normalization condition for the state  $|\mu>$
leads to
\begin{equation}
1=<\mu|\mu>=|{\cal{N}}|^2\sum_{n=0}^\infty
\frac{(\mu^*\mu)^n}{n!}\frac{\Gamma(2j+k)}{\Gamma(2j+k+n)}
\end{equation}
That is to say
\begin{equation}
|{\cal{N}}|^{-2}=\Gamma(2j+k)|\mu|^{-(2j+k-1)}I_{2j+k-1}(2|\mu|)
\end{equation}
where $I_\nu (x)$ is the modified Bessel function \cite{bate} and
$|\mu|^2=\mu^*\mu$. To ensure that the right hand side of eq.(2.23) is
positive for
arbitrary complex $\mu$, we take $2j+k$ positive. Thus we get the spin-level
restriction [4-6,8-14]
\begin{equation}
j>-\frac{k}{2}
\end{equation}
Therefore,  altogether, for $2j+k$ positive, we have
\begin{eqnarray}
|\mu>&=&{\cal{N}}\sum_{n=0}^\infty\mu^n
\frac{\Gamma(2j+k)}{\Gamma(n+1)\Gamma(2j+k+n)}\left( J^+_{-1}\right)^n
|jj>\nonumber\\
&=&{\cal{N}}\;\Gamma(2j+k)\left(\mu
J^+_{-1}\right)^{-(2j+k-1)/2}I_{2j+k-1}\left(2\sqrt{\mu J^+_{-1}}\right)|jj>
\end{eqnarray}
where  ${\cal{N}}$ is given by eq.(2.23).
As in quantum mechanics, the coherent states (2.25) do not form an
orthogonal set.
The scalar product of two coherent states is given by
\begin{equation}
|<\nu|\mu>|^2=\frac{I_{2j+k-1}(2\sqrt{\nu^*\mu})I_{2j+k-1}(2\sqrt{\mu^*\nu})}
{I_{2j+k-1}(2\sqrt{\nu^*\nu})I_{2j+k-1}(2\sqrt{\mu^*\mu})}
\end{equation}

In string theory, a physical state  must fulfill the mass-shell condition
and the
Virasoro primary conditions
\begin{equation}
(L_0-1)|\psi>=0,\;\;\;\;\;\;L_l|\psi>=0;\;\;\;l>0
\end{equation}
For the coherent states (2.25) it is easy to see that the Virasoro
primary conditions are fulfilled. However, being coherent states, they
obviously are not eigenstates of neither the number operator nor of the $L_0$
operator. We shall therefore impose a "weak" mass-shell condition
\begin{equation}
<\mu|( L_0 -1)|\mu>=0
\end{equation}
Using the identity
\begin{equation}
L_0\left(J^+_{-1}\right)^n|jj>=\left(n-\frac{j(j+1)}{k-2}\right)
\left(J^+_{-1}\right)^n|jj>
\end{equation}
the condition (2.28) leads to
\begin{equation}
\frac{I_{2j+k}(2|\mu|)}{I_{2j+k-1}(2|\mu|)}=|\mu|^{-1}
\left(1+\frac{j(j+1)}{k-2}\right)
\end{equation}
which is to be solved for (say) $j$ as a function of $\mu$.

\bigskip

We close this section with some comments on the case where
$2j+k=\{0,-1,-2,...\}=-N$, where $N$ is a non-negative integer. In that
case the
solution (2.20) for $ C_n $ is actually not well-defined. Instead the recursion
relation (2.19)
is solved by
\begin{equation}
\left\{ \begin{array}{ll}
 C_n=0;\;\;\;\;\;\;n=0,1,...N\\
C_{N+1+l}=\frac{\mu^l}{l!}\frac{\Gamma(N+2)}{\Gamma(N+2+l)};\;\;\;\;\;\;l\geq 0
\end{array}\right.
\end{equation}
and the coherent state is given by
\begin{eqnarray}
|\mu>&=&{\cal{N}}\sum_{n=0}^\infty\mu^n
\frac{\Gamma(N+2)}{\Gamma(n+1)\Gamma(N+2+n)}\left( J^+_{-1}\right)^{N+1+n}
|-(N+k)/2, -(N+k)/2>\nonumber\\
&=&{\cal{N}}\;\frac{\Gamma(N+2)}{\mu^{N+1}}\left(\mu
J^+_{-1}\right)^{(N+1)/2}I_{N+1}\left(2\sqrt{\mu J^+_{-1}}\right)|-(N+k)/2,
-(N+k)/2>\nonumber\\
\end{eqnarray}
where  ${\cal{N}}$ is a normalization constant. However, using the identity
\begin{eqnarray}
<-(N+k)/2, -(N+k)/2|\left(J^-_{+1}\right)^n
\left(J^+_{-1}\right)^m|-(N+k)/2,
-(N+k)/2>\nonumber\\=(-1)^n\delta_{nm}n!\left\{
\begin{array}{cc}\frac{N!}{(N-n)!};\;\;\;n\leq N\\
0;\;\;\;n>N\end{array}\right.\nonumber\\
\end{eqnarray}
one finds that $<\mu|\mu>=0$. That is to say, the coherent states for
$2j+k=-N$ are
zero norm states, so we shall not consider them further.

We are thus left with a continous spectrum of the
positive norm coherent states (2.25). The discrete spectrum of states
(2.32) are
all zero norm states, and are expected to decouple in scattering amplitudes.
\section{Minimal Uncertainty}
\setcounter{equation}{0}
One of the most important properties of coherent states in quantum
mechanics is the
one of minimal uncertainty \cite{merz}
\begin{equation}
\Delta X\cdot \Delta P=\frac{1}{2}
\end{equation}
We shall now show that the states (2.25) lead to the same property in the
case of a
Kac-Moody algebra. First we define Hermitean operators $(X,P,{\cal H})$
\begin{eqnarray}
X&\equiv&\frac{1}{\sqrt{2k}}(J^-_{+1}+J^+_{-1})\nonumber\\
P&\equiv&\frac{-i}{\sqrt{2k}}(J^-_{+1}-J^+_{-1})\\
{\cal H}&\equiv&J^3_0\nonumber
\end{eqnarray}
Then, the algebra (2.13) becomes
\begin{eqnarray}
&[X,P]&= i(1+\frac{2}{k}{\cal H})\nonumber\\
&[X,{\cal H}]&=iP\\
&[{\cal H},P]&=iX\nonumber
\end{eqnarray}
The algebra (3.3) can be interpreted as a modified Harmonic oscillator
algebra; the
modification being represented by the second term $ \frac{2}{k}{\cal
H} $ in the $ X, \; P $ commutator. That is,
in the semi-classical limit $(k\rightarrow\infty)$, we get the standard
Harmonic
oscillator algebra \cite{merz}. It is quite natural to interprete $X$ and
$P$ as coordinate and momentum, respectively, since, in the context of
Kac-Moody algebras, the roles of
coordinates and momenta are played by the currents $J$. However, the
interpretation of ${\cal H}$ as some kind of Hamiltonian needs a few
comments: First, notice that,
contrary to the case of the standard harmonic oscillator, ${\cal
H}$ is here an {\it independent} operator; in particular, ${\cal
H}\neq\frac{1}{2}(P^2+X^2)$. On the other hand, there is a simple relation
between
$J^3_0$ and the energy $E$ and angular momentum $l$ of a string in $AdS_3$
\cite{all,san2,Ooguri}
\begin{eqnarray}
J^3_0&=&\frac{1}{2\pi}\int_0^{2\pi}J^3\;d\sigma=\frac{1}{2}(E+l)\nonumber\\
\bar{J}^3_0&=&\frac{1}{2\pi}\int_0^{2\pi}\bar{J}^3\;d\sigma=\frac{1}{2}(E-l)
\end{eqnarray}
where a bar denotes right-movers. Thus, the total energy is
$E=J^3_0+\bar{J}^3_0$ and
it is natural to identify ${\cal H}\sim J^3_0$.

From eqs.(3.3), the  uncertainty relation here is given by
\begin{equation}
\Delta X\cdot \Delta P\geq\frac{1}{2}|<(1+\frac{2}{k}{\cal H})>|
\end{equation}
with minimal uncertainty in the case of equality sign. For $ k \to
\infty $, it is the usual Heisenberg relation.

Now, consider the coherent states (2.25). It is straightforward to compute
\begin{equation}
(\Delta X)^2=(\Delta P)^2=\frac{1}{2k}\left(k+2(j+1)(1+\frac{j}{k-2})\right)
\end{equation}
as well as
\begin{equation}
<{\cal H}>=(j+1)(1+\frac{j}{k-2})
\end{equation}
That is to say
\begin{equation}
\Delta X\cdot \Delta P=\frac{1}{2}|<(1+\frac{2}{k}{\cal H})>|
\end{equation}
i.e., minimal uncertainty. That is to say, states (2.25) are quasiclassical
states.

\section{"Exponential" Coherent States}
\setcounter{equation}{0}

In a previous paper \cite{all}, we considered a different type of
coherent
states defined in terms of the exponential of the creation operator
\begin{equation}
e^{\tilde{\mu} J^+_{-1}}|jj>
\end{equation}
where $\tilde{\mu}$ is an arbitrary
complex number. Such coherent states (4.1) are however not eigenstates of the
annihilation operator $J^-_{+1}$
\begin{equation}
J^-_{+1} e^{\mu J^+_{-1}}|jj>=\mu\left( 2j+k+\mu J^+_{-1}\right)
e^{\mu J^+_{-1}}|jj>
\end{equation}
As for the normalization of states (4.1), we use the identity
\begin{equation}
<jj|e^{\tilde{\mu}^* J^-_{+1}}\; e^{\tilde{\mu} J^+_{-1}}|jj>=1+
\sum_{n=1}^\infty\frac{(\tilde{\mu}^*\tilde{\mu})^n}{n!}\prod_{l=1}^n
(2j+k-1+l)
\end{equation}
The product on the right hand side goes as $n!$. Thus the infinite sum is
convergent
only if $\tilde{\mu}^*\tilde{\mu}<1$, or if the infinite sum terminates after a
finite number of terms (this happens if $2j+k-1+l=0$, for some $l$).
More precisely, the right hand side of eq.(4.3) is a finite positive
number in the following two cases\\
\\
{\bf (I):} $\tilde{\mu}^*\tilde{\mu}<1$ and $j$ arbitrary ($j\leq -1/2$).\\
In this case the normalized state is
\begin{equation}
|\tilde{\mu}_I>=(1-\tilde{\mu}^*\tilde{\mu})^{j+k/2}\;e^{\tilde{\mu}
J^+_{-1}}|jj>
\end{equation}
\\
{\bf (II):} $\tilde{\mu}^*\tilde{\mu}>1$ and
$2j+k=-N\;\;\;\;(N={0,1,2,...}).$\\ In this case the normalized state is
\begin{equation}
|\tilde{\mu}_{II}>=(\tilde{\mu}^*\tilde{\mu}-1)^{-N}\;e^{\tilde{\mu}
J^+_{-1}}|-N-k/2, -N-k/2>
\end{equation}
The Virasoro primary conditions are obviously fulfilled for the states
(4.4)-(4.5),
while the mass-shell condition in the form of eq.(2.28) gives rise to some
additional constraints on
$\tilde{\mu}$ and
$j$. In the two cases one finds, respectively\\
\\
{\bf (I):}\\
\begin{equation}
\tilde{\mu}^*\tilde{\mu}=\frac{1+\frac{j(j+1)}{k-2}}{2j+k+1+\frac{j(j+1)}{k-2}}<
1\;
;\;\;\;\;
\;\;\;\;-\frac{k}{2}<j\leq -\frac{1}{2}
\end{equation}\\
\\
{\bf (II):}\\
\begin{equation}
\tilde{\mu}^*\tilde{\mu}=\frac{1+\frac{j(j+1)}{k-2}}{2j+k+1+\frac{j(j+1)}{k-2}}>
1\;
;\;\;\;\;
\;\;\;\; j=-N-\frac{k}{2}\;\;\;\;(N={1,2,...})
\end{equation}
It follows that the spectrum consists of two parts \cite{all}: {\bf (I)}
A continuous spectrum where $j$
fulfills the standard spin-level condition
[4-6,8-14]
$-k/2 <j\leq -1/2$, and {\bf (II)}
a discrete spectrum where $j$ fulfills  $j=-N-k/2\;$ ($N$ positive
integer). The discrete spectrum describes  very
massive string states, with asymptotic behaviour $m^2\alpha'\propto N^2$ ($N$
positive integer) \cite{all}. This is in precise
agreement with previous results obtained using semi-classical
quantization \cite{san2,vega1,vega}, and the same asymptotic behaviour was also
obtained in the recent paper \cite{Ooguri}.

Unfortunately, the states (4.4)-(4.5) are somewhat complicated to work with
since
they are not eigenstates of the annihilation operator. Therefore it would be
useful to express them in terms of the states (2.25). More generally, let us
consider the off-shell relationship between states (2.25) and states (4.4).
Clearly,
the two types of coherent states are not orthogonal
\begin{equation}
<\mu|\tilde{\mu}_I>=\frac{(1-\tilde{\mu}^*\tilde{\mu})^{j+k/2}\;
|\mu|^{(2j+k-1)/2}\;e^{\mu^*\tilde{\mu}}}{\sqrt{\Gamma(2j+k)I_{2j+k-1}(2|\mu|)}}
\end{equation}
but it is possible to express (say) the states $|\tilde{\mu}_I>$ in terms
of the
states $|\mu>$. However, since the states $|\mu>$ form an over-complete
set, the
expression will of course not be unique. So, we just give an example:

For fixed $2j+k$, we first introduce normalized basis states $|n>$
\begin{equation}
|n>=\sqrt{\frac{\Gamma(2j+k)}{\Gamma(n+1)\Gamma(2j+k+n)}}\;\left(
J^+_{-1}\right)^n|jj>;\;\;\;\;\;<n|m>=\delta_{nm}
\end{equation}
It follows that
\begin{equation}
<n|\mu>=\frac{|\mu|^{(2j+k-1)/2}\mu^n}
{\sqrt{\Gamma(n+1)\Gamma(2j+k+n)I_{2j+k-1}(2|\mu|)}}
\end{equation}
as well as
\begin{equation}
|n>=\int d^2\mu\; f_n(\mu)|\mu>
\end{equation}
where
\begin{equation}
f_n(\mu)=\frac{{\mu^*}^n
e^{-\frac{1}{2}\mu^*\mu}}{2^{n+1}\pi|\mu|^{(2j+k-1)/2}}
\sqrt{\frac{\Gamma(2j+k+n)}{\Gamma(n+1)}I_{2j+k-1}(2|\mu|)}
\end{equation}
Using  eq.(4.4), we eventually get the formal expression\begin{equation}
|\tilde{\mu}_I>=(1-\tilde{\mu}^*\tilde{\mu})^{j+k/2}\sum_{n=0}\tilde{\mu}^n
\sqrt{\frac{\Gamma(2j+k+n)}{\Gamma(n+1)\Gamma(2j+k)}}\;
\int d^2\mu\; f_n(\mu)|\mu>
\end{equation}
which is the desired result.

\section{Conclusion}

We studied  the properties of exact (all level $k$) quantum coherent  states in
the context of  Kac -Moody algebras (WZWN models).

Quantum coherent states in the context of string theory on a group manifold
(WZWN models) are important since they may help solving the unitarity problem:
Having positive norm (no ghost-states appear in the string spectrum), they
consistently include the high massive strings (which otherwise are excluded
by the
spin-level condition).

Coherent  states admit (at least) two alternative definitions: (i) as the
exponential of the creation operator acting on the ground state, and (ii) as
eigenstates of the annhilation operator. In the $ k \to \infty $
limit, all the known properties of usual coherent states are recovered.

In ordinary quantum mechanics and string theory in flat space time (with the
usual commutator algebra of harmonic oscillators), the two alternative
definitions
(i) and (ii) are equivalent. This is not the case in the context of Kac-Moody
algebras  as in the WZWN  models.

In this paper we have constructed  coherent states as defined by (ii), compared
them to the states (i) we previously constructed, and computed the uncertainty
relation in this context. Modification to the Heisenberg relation is given by
$ 2\,{\cal H }/k $ where  ${\cal H }$ is connected to the string
energy. Coherent states (ii) are
generally much easier to work with and satisfy automatically and for any
$k$ the
minimal uncertainty condition. They are thus quassiclassical, in some
sense more classical than states (i) which only satisfy it in the
$ k \to \infty $ limit.

The coherent states (ii) reduce, for $k=0$, to the group-related $SU(1,1)$
coherent
states constructed in \cite{barut}, as explained in the introduction. In the
opposite limit, for $k \rightarrow \infty$, they reduce to the standard
canonical
coherent states \cite{merz}.

\vskip 12pt
{\bf Acknowledgements}\\
A.L.L. would like to thank the {\bf Ambassade de France \`a Copenhague},
{\bf Service Culturel et Scientifique}  for financial support in Paris,
during the preparation of this paper.


\newpage
\begin{thebibliography}{11}
\bibitem{merz}E. Merzbacher, "Quantum Mechanics" (Wiley, New York, 1970).
\bibitem{green}M.B. Green, J.H. Schwarz and E. Witten, "Superstring Theory"
(Cambridge University Press, Cambridge, 1994).
\bibitem{balog}J. Balog, L. O'Raifeartaigh, P. Forgacs and A. Wipf,
Nucl.
Phys. {\bf B325} (1989) 225.
\bibitem{dixon}L.J. Dixon, M.E. Peskin and J. Lykken, Nucl. Phys. {\bf B325}
(1989) 329.
\bibitem{petr}P. Petropoulos, Phys. Lett. {\bf B236} (1990) 151.
\bibitem{moh}N. Mohammedi, Int. Journ. Mod. Phys. {\bf A5} (1990) 3201.
\bibitem{nem}I. Bars and D. Nemeschansky, Nucl. Phys. {\bf B348}
(1991) 89.
\bibitem{bars}I. Bars, Nucl. Phys. {\bf B334} (1990) 125.
\bibitem{hwa}S. Hwang, Nucl. Phys. {\bf B354} (1991) 100.
\bibitem{maans1}M. Henningson and S. Hwang, Phys. Lett. {\bf B258}
(1991) 341.
\bibitem{maans2}M. Henningson, S. Hwang, P. Roberts and B. Sundborg, Phys.
Lett.
{\bf B267} (1991) 350.
\bibitem{hwang}S. Hwang, Phys. Lett. {\bf B276} (1992) 451.
\bibitem{evans}J.M. Evans, M.R. Gaberdiel and M.J. Perry, Nucl. Phys.
{\bf B535} (1998) 152.
\bibitem{satoh}M. Natsuume and Y. Satoh, Int. J. Mod. Phys. {\bf A13}
(1998) 1229.
\bibitem{ibars}I. Bars, Phys. Rev. {\bf D53} (1996) 3308.
\bibitem{ysatoh}Y. Satoh, Nucl. Phys. {\bf B513} (1998) 213.
\bibitem{malda}J. Maldacena, Adv. Theor. Math. Phys. {\bf 2} (1998) 231.
\bibitem{sen}M. Mathur and D. Sen, {\it Coherent States For $SU(3)$}. Preprint,
quant-ph/0012099.
\bibitem{all}A.L. Larsen and N. S\'{a}nchez, Phys. Rev. {\bf D62}
(2000) 046003.
\bibitem{san2}A.L. Larsen and N. S\'{a}nchez, Phys. Rev. {\bf D52} (1995)
1051.
\bibitem{vega1}H.J. de Vega, A.L. Larsen and N. S\'{a}nchez,
Phys. Rev. {\bf D51} (1995) 6917.
\bibitem{vega} H.J. de Vega, A.L. Larsen and  N. S\'{a}nchez,
 Phys. Rev.
{\bf D58} (1998) 026001.
\bibitem{Ooguri}J. Maldacena and H. Ooguri, {\it Strings in $AdS_3$ and
$SL(2,R)$ WZW Model: I}. Preprint, hep-th/0001053.
\bibitem{klauder}J.R. Klauder and B.S. Skagerstam, "Coherent States" (World
Scientific, Singapore, 1985).
\bibitem{barut}A.O. Barut and L. Girardello, Commun. Math. Phys. {\bf 21}
(1971) 41.
\bibitem{wit}E. Witten, Commun. Math. Phys. {\bf 92} (1984) 455.
\bibitem{wit2}D. Gepner and E.Witten, Nucl. Phys. {\bf B278} (1986) 493.
\bibitem{barg}V. Bargmann, Ann. Math. {\bf 48} (1947) 568.
\bibitem{bate}H Bateman, "Higher Transcendental Functions" (McGraw-Hill,
New York,
1955) Vol II.
\end{thebibliography}
\end{document}